\documentclass[twocolumn,hyperpdf,amsmath,amssymb,aps,prl,10pt,superscriptaddress,nofootinbib,noeprint,preprintnumbers]{revtex4-1}

\usepackage{graphicx}
\usepackage{dcolumn}
\usepackage{bm}
\usepackage{amsmath}
\usepackage{amsfonts}
\usepackage{slashed}
\usepackage{hyperref}
\usepackage{multirow}
\usepackage{soul}
\usepackage{mathtools}

\usepackage{afterpage}

\usepackage{color}

\hypersetup{
pdftitle = {Resonances in coupled $\pi K, \eta K$ scattering from Quantum Chromodynamics},
pdfsubject = {QCD},
pdfkeywords = {QCD, Lattice, scattering, hadron, spectrum, Kaon, pion, eta},
pdfauthor = {Jozef J. Dudek, Robert G. Edwards, Christopher E. Thomas, David J. Wilson},
colorlinks = {false},
}{}

\begin{document}

\preprint{JLAB-THY-14-1891}
\preprint{DAMTP-2014-36}

\title{Resonances in coupled $\pi K, \eta K$ scattering from quantum chromodynamics}

 \author{Jozef~J.~Dudek}
  \email{dudek@jlab.org}
 \affiliation{Theory Center, Jefferson Lab, 12000 Jefferson Avenue, Newport News, VA 23606, USA}
 \affiliation{Department of Physics, Old Dominion University, Norfolk, VA 23529, USA}

 \author{Robert~G.~Edwards}
 \affiliation{Theory Center, Jefferson Lab, 12000 Jefferson Avenue, Newport News, VA 23606, USA}

 \author{Christopher~E.~Thomas}
 \affiliation{DAMTP, University of Cambridge, Cambridge, UK}

 \author{David~J.~Wilson}

 \affiliation{Department of Physics, Old Dominion University, Norfolk, VA 23529, USA}

\collaboration{for the Hadron Spectrum Collaboration}

\date{\today}

\begin{abstract}
Using first-principles calculation within Quantum Chromodynamics, we are able to determine a pattern of strangeness=1 resonances which appear as complex singularities within coupled $\pi K$, $\eta K$ scattering amplitudes. We make use of numerical computation in the lattice discretized approach to the quantum field theory with light quark masses corresponding to $m_\pi \sim 400\, \textrm{MeV}$ and at a single lattice spacing. The energy dependence of scattering amplitudes is extracted through their relationship to the discrete spectrum in a finite-volume, which we map out in unprecedented detail.
\end{abstract}


\maketitle

\vspace{-1cm}

\paragraph{Introduction:}

Understanding how excited hadrons are built up from the basic quark and gluon degrees-of-freedom of Quantum Chromodynamics (QCD) remains a challenging problem. QCD should be able to explain the apparent regularity of the experimental spectrum, but its structure suggests additional ``exotic" quark-gluon configurations, such as those in which the gluonic field is excited, which to date have not been unambiguously observed in experiment. An important feature of the theoretical challenge is that excited hadrons appear in experiment as \emph{resonances}, which ultimately decay into pseudoscalar mesons, themselves complex aggregations of confined quarks, antiquarks and gluons. To study resonances within QCD, one must explore the behavior of the theory's hadron scattering amplitudes at low energy, a field of study that is still in its infancy.

Lattice QCD offers us an \emph{ab initio} numerical approach to QCD calculations; by discretizing the quark and gluon fields on a finite lattice, and Monte-Carlo sampling possible space-time configurations of the gluon field, we can compute correlation functions with the quantum numbers of hadrons. These correlators contain information about the spectrum and interactions of hadrons.

Following L\"uscher~\cite{Luscher:1990ux}, a formalism has been derived which relates infinite-volume hadron scattering amplitudes to the discrete spectrum of hadron states in a finite-volume. Through computation of statistically precise excited state spectra in lattice QCD we can extract the energy dependence of hadron scattering amplitudes and examine their resonant content. In a recent example considering the elastic scattering of two pions~\cite{Dudek:2012xn}, following computation of the appropriate lattice QCD spectra, the scattering amplitude was determined at 32 discrete energy values in a 300 MeV range, demonstrating unambiguously the presence of the $\rho$ resonance, whose mass and width were extracted. 

Unlike the $\rho$ resonance, most known hadron resonances do not decay into only one final state, rather being enhancements in \emph{coupled-channels}; the recently derived formalism to extract coupled-channel scattering amplitudes from finite-volume spectra is somewhat more involved than the elastic case~\cite{Guo:2012hv, He:2005ey, *Hansen:2012tf,*Briceno:2012yi}. In this letter we will present the first application of this formalism to QCD, for the particular case of $\pi K, \eta K$ coupled-channel scattering.

Experimentally a number of low-lying resonances appear in isospin--1/2 $\pi K$ scattering: the $J^P=1^-$, $K^\star(892)$, the strange analogue of the $\rho$ meson, is a narrow elastic resonance, the $J^P=0^+$, $K^\star_0(1430)$, appears as a relatively broad resonance and the $J^P=2^+$, $K_2^\star(1430)$, at a similar mass value, is a much narrower resonance~\cite{Beringer:1900zz}. Further, it is observed that in \mbox{even-$J$} partial-waves, resonances decay dominantly to $\pi K$ and not to $\eta K$.

Resonances can be rigorously defined as pole singularities in scattering amplitudes when they are considered to be functions of \emph{complex} values of the scattering energy -- experimentally the scattering amplitudes are determined for real values of the energy above kinematic thresholds, which may then be parameterized by analytic functions that can be continued into the complex-plane. One particularly important application of this procedure concerns the $J^P=0^+$ $\pi K$ scattering amplitude at low energy, where the use of dispersion relations provides a particularly strong constraint when describing the available experimental scattering data, leading to a pole far from the real axis known as the $\kappa$~\cite{DescotesGenon:2006uk}.

In this letter we will present scattering amplitudes extracted from lattice QCD determinations of the excited hadron spectrum for the $J^P=0^+,1^-, 2^+$ coupled $\pi K, \eta K$ channels. In this first calculation, we work with light quark masses in the QCD Lagrangian somewhat heavier than those known physically, meaning that initially our results can only be compared qualitatively with the experimental situation. We find, as in experiment, an approximate decoupling of the $\pi K, \eta K$ channels and a rapid increase of the $\pi K$ $J^P=0^+$ scattering phase-shift above threshold, followed by a slower rise through $90^\circ$ at higher energies. With light quark masses somewhat heavier than the physical case, the lightest vector resonance becomes bound and the $\kappa$ pole appear to manifest itself as a ``virtual bound state". Behavior consistent with a narrow resonance is observed in the $J^P=2^+$ channel.

Previous lattice QCD studies limited to elastic $\pi K$ scattering in isospin--1/2 appear in Refs. \cite{Fu:2011wc, *Fu:2012tj, *Sasaki:2013vxa, Lang:2012sv, *Prelovsek:2013ela}. Refs.~\cite{Lang:2012sv, *Prelovsek:2013ela}, in calculations in a small volume with lower light-quark masses compared to this study and without dynamical strange quarks, found a $K^\star$ above $\pi K$ threshold and extracted resonance parameters.


\paragraph{Finite-volume spectrum from lattice QCD:}

\begin{figure*}
\includegraphics[width=0.85\textwidth]{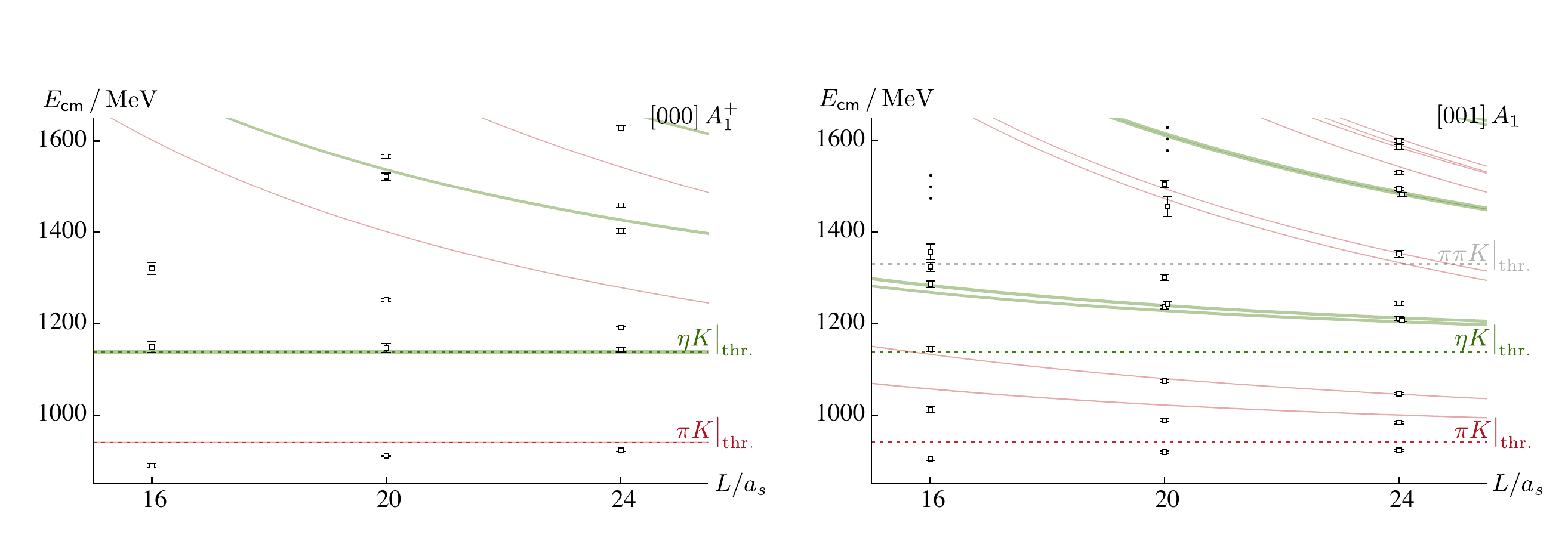}
\caption{Discrete spectrum of $\mathsf{cm}$-frame energies obtained from variational analysis of correlation matrices featuring ``single-meson" and ``meson-meson" operators at three spatial volumes. Red bands are $\pi K$ non--interacting level positions, green bands are $\eta K$ non--interacting level positions. 
(a) $\vec{P}=[000]$, $\Lambda = A_1^+$ spectrum (dominated by $J^P=0^+$ with negligible contributions from $J\ge 4$). (b) $\vec{P}=[001]$, $\Lambda = A_1$ spectrum ($J^P=0^+, 1^-, 2^+$ all contribute).}
\label{spectra}
\end{figure*}

We make use of the dynamical anisotropic lattices initially described in~\cite{Lin:2008pr}, which feature two degenerate light quarks ($u,d$) plus a heavier flavor with its mass tuned to approximate the physical strange quark. In this study the light quark mass parameter is increased with respect to its physical value such that the pion mass is 391 MeV, the kaon mass is 549 MeV and the $\eta$ mass is 589 MeV. Three spatial volumes are utilized: ${16^3(\sim \!2\,\mathrm{fm}),\, 20^3(\sim\! 2.5\,\mathrm{fm}), \,24^3(\sim\! 3\,\mathrm{fm})}$, with a spatial lattice spacing, $a_s \sim 0.12 \,\mathrm{fm}$.

The discrete spectrum of hadron states can be extracted from two-point correlation functions, $\big\langle 0 \big| \mathcal{O}^{}_i(t) \, \mathcal{O}^\dag_j(0) \big| 0 \big\rangle$. We make use of two classes of interpolating field, $\mathcal{O}^\dag_i$; the first are ``single-meson" operators, those which resemble a $q\bar{q}$ construction, $\bar{\psi}\mathbf{\Gamma} \psi$, where $\mathbf{\Gamma}$ is one of a large number of operators in spin, color and position space~\cite{Dudek:2009qf, *Dudek:2010wm, *Thomas:2011rh}, while the second resembles a pair of mesons with definite relative and total momentum, $\big(\bar{\psi}\mathbf{\Gamma}_1 \psi\big)_{\vec{p}_1}\, \big(\bar{\psi}\mathbf{\Gamma}_2 \psi\big)_{\vec{p}_2}$~\cite{Dudek:2012gj}. In an $L\!\times\! L \!\times\! L$ spatial volume with periodic boundary conditions, the momenta which appear in these constructions are quantized in integer multiples of $\frac{2\pi}{L}$: $\vec{p} = \frac{2\pi}{L}(n_x, n_y, n_z)$. More details of the constructions and a demonstration of their efficacy can be found in~\cite{Dudek:2012gj}.

For this study we construct a basis of operators including ``meson-meson" constructions with quark flavors corresponding to $\pi K$ and $\eta K$. A matrix of correlation functions is built~\cite{Peardon:2009gh} using a large number of ``single-meson" operators along with several relative momentum constructions of both $\pi K$-like and $\eta K$-like operators. A spectrum of states that is best in a variational sense can be extracted from this matrix by solving a generalized eigenvalue problem~\cite{Michael:1985ne,*Luscher:1990ck}.

The lattice has a cubic symmetry, rather than that of the full rotational group, so instead of being characterized by integer spins, independent spectra lie in a finite number of irreducible representations, or \emph{irreps}, $\Lambda$, of the reduced symmetry. For a meson system with non-zero total momentum, $\vec{P}$, the symmetry group is classified by irreps of rotations that leave $\vec{P}$ invariant~\cite{Thomas:2011rh}. We compute the excited state spectrum in all relevant irreps for $|\vec{P}|^2 \le 4 \left( \tfrac{2\pi}{L} \right)^2$ and find we can determine them with a high degree of statistical precision in all cases. In Fig.~\ref{spectra} we show two examples -- note that owing to the use of the variational method of state extraction, there is robust extraction of near-degenerate states, even high in the spectrum.

If mesons had no residual strong interactions, the discrete spectrum in a finite-volume would be predictable using the relativistic dispersion relation ${E = \sqrt{m_1^2 + |\vec{p}_1|^2} +  \sqrt{m_2^2 + |\vec{p}_2|^2} }$ with the allowed quantized momentum values given above. These volume-dependent ``non-interacting" energy levels are shown by the continuous curves in Fig.~\ref{spectra}, and the fact that the energy levels extracted from our lattice QCD calculations do not lie on these curves is an indication that there are interactions between mesons. If these interactions are strong enough, there may be resonant behavior; to explore this rigorously, we must determine the meson-meson scattering amplitudes.


\paragraph{Scattering amplitudes:}

\begin{figure*}
\makebox[\textwidth][c]{\includegraphics[width=1.1\textwidth]{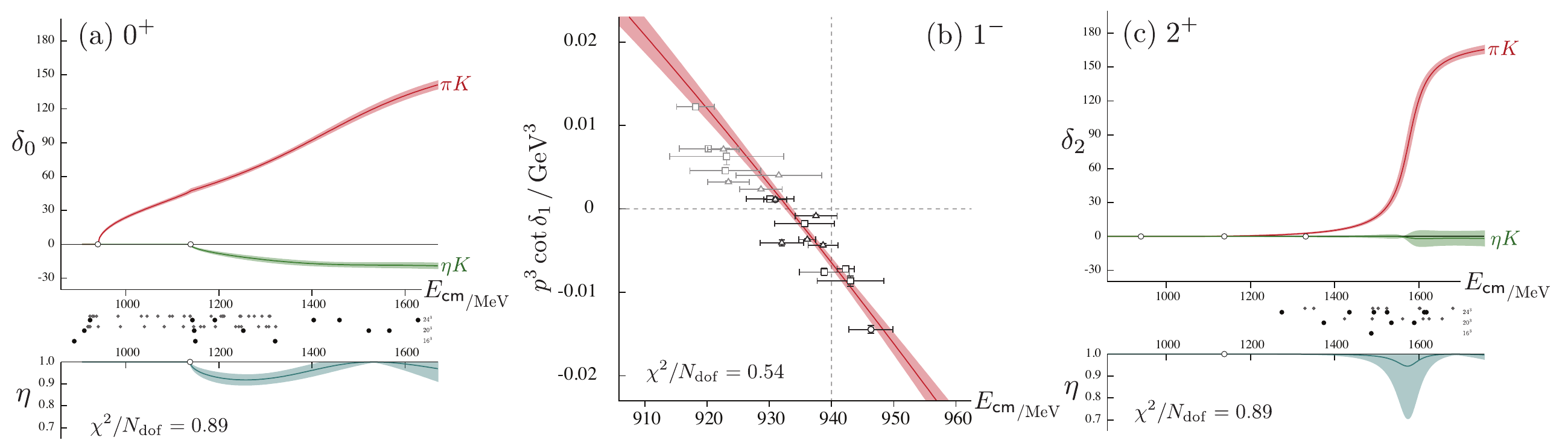}}
\caption{(a) $J^P=0^+$ amplitudes -- open circles on axis show $\pi K$ and $\eta K$ thresholds. Upper panel: $\pi K$ and $\eta K$ phase-shifts in degrees. Lower panel: inelasticity. Points in center show the energy levels on three volumes used to constrain the $t$-matrix extraction -- larger solid points show $\vec{P}=\vec{0}$, smaller open circles show $\vec{P} \neq\vec{0}$. (b) $J^P=1^-$ around the $\pi K$ threshold. Points determined directly without parameterization of the vector amplitude from three volumes: $16^3$(boxes), $20^3$(circles) and $24^3$(triangles). Curve shows the result of a relativistic Breit-Wigner parameterization, ${p^3 \cot \delta_1 = (m_R^2 -s) \tfrac{6\pi \sqrt{s}}{g_R^2}  }$. (c) $J^P=2^+$ amplitudes -- open circles on axis show $\pi K$, $\eta K$ and $\pi\pi K$ thresholds. }
\label{master}
\end{figure*}

For an interacting quantum field theory, the relationship between the discrete spectrum of states in an $L\times L \times L$ volume, in a frame moving with momentum $\vec{P}$, for irrep $\Lambda$, can be written in a simplified form as
\begin{equation}
\det\Big[
\delta_{ij}\delta_{J J^\prime} + i\rho^{}_i\, t^{(J)}_{i j}(E_\mathsf{cm})
\left(\delta_{J J^\prime} +
i  \mathcal{M}^{\vec{P}\Lambda}_{J J^\prime}(p_i L) \right)
\Big] = 0, \nonumber
\label{eq_luescher_t}
\end{equation}
where the scattering $t$-matrix for partial-wave $J$ appears along with the phase-space in channel $i$, $\rho_i(s) = 2 p_i/E_\mathsf{cm}$, and the known volume-dependent functions, $\mathcal{M}$~\cite{Guo:2012hv, He:2005ey, *Hansen:2012tf,*Briceno:2012yi, Rummukainen:1995vs, *Leskovec:2012gb}.
Given knowledge of the energy-dependence of the scattering amplitudes, $t^{(J)}_{ij}(E_\mathsf{cm})$, one can solve this equation for a discrete spectrum of states, $\{E_\mathsf{cm}\}$. The practical problem at hand, however, is the reverse of this: to find the $t$-matrix given a lattice QCD calculation of the spectrum. For any single energy level value, $E_\mathsf{cm}$, this is an underconstrained problem as there are multiple elements of the $t$-matrix to be determined from only one condition.

The approach we will take is to parameterize the energy-dependence of the $t$-matrix and describe the spectrum as a whole. Such an approach was explored in the context of a toy-model of coupled-channel scattering in \cite{Guo:2012hv}. A flexible $K$-matrix parameterization of partial-wave $J$, in terms of the variable $s=E_\mathsf{cm}^2$, can be constructed which ensures the unitarity of the $S$-matrix,
\begin{align}
	t^{-1}_{ij}(s) &= \frac{1}{(2p_i)^J} K^{-1}_{ij}(s) \frac{1}{(2p_j)^J} + I_{ij}(s), \nonumber \\
	K_{ij}(s) &= \sum_p\frac{g^{(p)}_{i} g^{(p)}_{j}}{m_p^2-s} + \sum_n\gamma_{ij}^{(n)}s^n , \nonumber
\label{k-matrix}
\end{align}
where we may choose how many poles and what order polynomial to include in $K$, with real parameters $g^{(p)}_i,\, m_p, \gamma^{(n)}_{ij}$. The function $I(s)$ must be chosen such that $\mathrm{Im}\, I_{ij}(s) = - \delta_{ij}\, \rho_i(s)$ above threshold to ensure unitarity is preserved. There is some freedom in the choice of the real part; we choose an implementation of the Chew-Mandelstam form~\cite{Guo:2012hv} which has smooth behavior across kinematic thresholds.

We use 80 levels from 20 irreps to constrain the $0^+$ amplitudes from slightly below $\pi K$ threshold up to 1650 MeV, 19 levels to constrain the $1^-$ amplitude in the region around the $\pi K$ threshold and a further 24 levels to constrain the $2^+$ amplitudes between 1250 and 1700 MeV. The $0^+, 2^+$ partial-waves are described by a single $K$-matrix pole coupled to both $\pi K$ and $\eta K$ plus a constant matrix, while the $\pi K$ threshold region in $1^-$ is described by a relativistic Breit-Wigner. We assume that the influence of partial-waves, $J \ge 3$, is negligible in this energy region.

The resulting $t$-matrices are plotted in Fig.~\ref{master} -- for $0^+,2^+$, $\pi K$ and $\eta K$ phase-shifts and an inelasticity, defined in $t_{ii}=\tfrac{(\eta e^{2i\delta_i}-1)}{2i \rho_i}$,
${t_{ij}= \tfrac{\sqrt{1-\eta^2} \,   e^{i(\delta_i+\delta_j)}}{ 2    \sqrt{\rho_i \, \rho_j}}}$, for channels $i=\pi K, \eta K$, are shown, while for $1^-$ we plot the function $p^3 \cot \delta$, which is real and continuous across the $\pi K$ threshold. In each case we present the $\chi^2/N_\mathrm{dof}$ for the parameterized description of the input spectrum, which we find to be quite acceptable.

The points shown in the center of Fig.~\ref{master}(a), which cover the whole energy region plotted, indicate that we are strongly constraining the energy dependence of the amplitudes; in particular note that the low-energy behavior of the $0^+$ $\pi K$ amplitude is constrained by points at or below threshold. Similarly in Fig.~\ref{master}(c), the energy dependence of the $2^+$ amplitude is well sampled in the region of the rapid rise of the phase-shift. This region is above the $\pi\pi K$ threshold, which can in principle couple to the $2^+$ partial-wave -- we have assumed here that there is negligible coupling to this channel.

We observe that the $\pi K$ phase-shifts in $0^+, 2^+$ rise through $90^\circ$ suggesting possible resonant behavior, and while this rise is rather slow in the scalar channel, it is rapid in the tensor, indicating a likely narrow resonance. The extracted inelasticities do not deviate significantly from unity corresponding to an approximate decoupling of the $\eta K$ channel from $\pi K$. The $\eta K$ amplitudes are found to be weak and repulsive.

The $1^-$ amplitude around the $\pi K$ threshold, shown in Fig.~\ref{master}(b), has the behavior expected of a bound-state, with the parameterization crossing zero below threshold. The energy dependence of the amplitude can be well described by a Breit-Wigner form, with a resulting mass of $m_R = 933(1)$ MeV, constrained by the zero-crossing, and coupling of $g_R = 5.93(26)$, constrained by the slope.


\paragraph{Resonance poles:} We may now analyze the singularity structure of our parameterized $t$-matrices by analytically continuing them to complex values of $s$. A general feature of relativistic scattering is that the square-root branch cut present in the phase-space leads to a multi-sheeted structure for scattering amplitudes. The usual approach is to place the branch cuts along the real axis beginning at kinematic thresholds and running out to positive infinity. Physical scattering occurs just above these cuts on the first Riemann sheet, sheet $\mathsf{I}$, where in our two-channel case, $\mathrm{Im}\, p_{\pi K} \!> \! 0$, $\mathrm{Im}\, p_{\eta K} \!>\! 0$. Resonance poles lie on the ``unphysical'' sheets reached by passing through the cuts, at pole positions which may be expressed in terms of a ``pole mass" and ``pole width", $\sqrt{s_0} = m - i \Gamma/2$. Bound-states may appear as poles on the physical sheet on the real axis below threshold, while poles on the real axis located on unphysical sheets are interpreted as ``virtual" bound-states.

\begin{figure}
\includegraphics[width=\columnwidth]{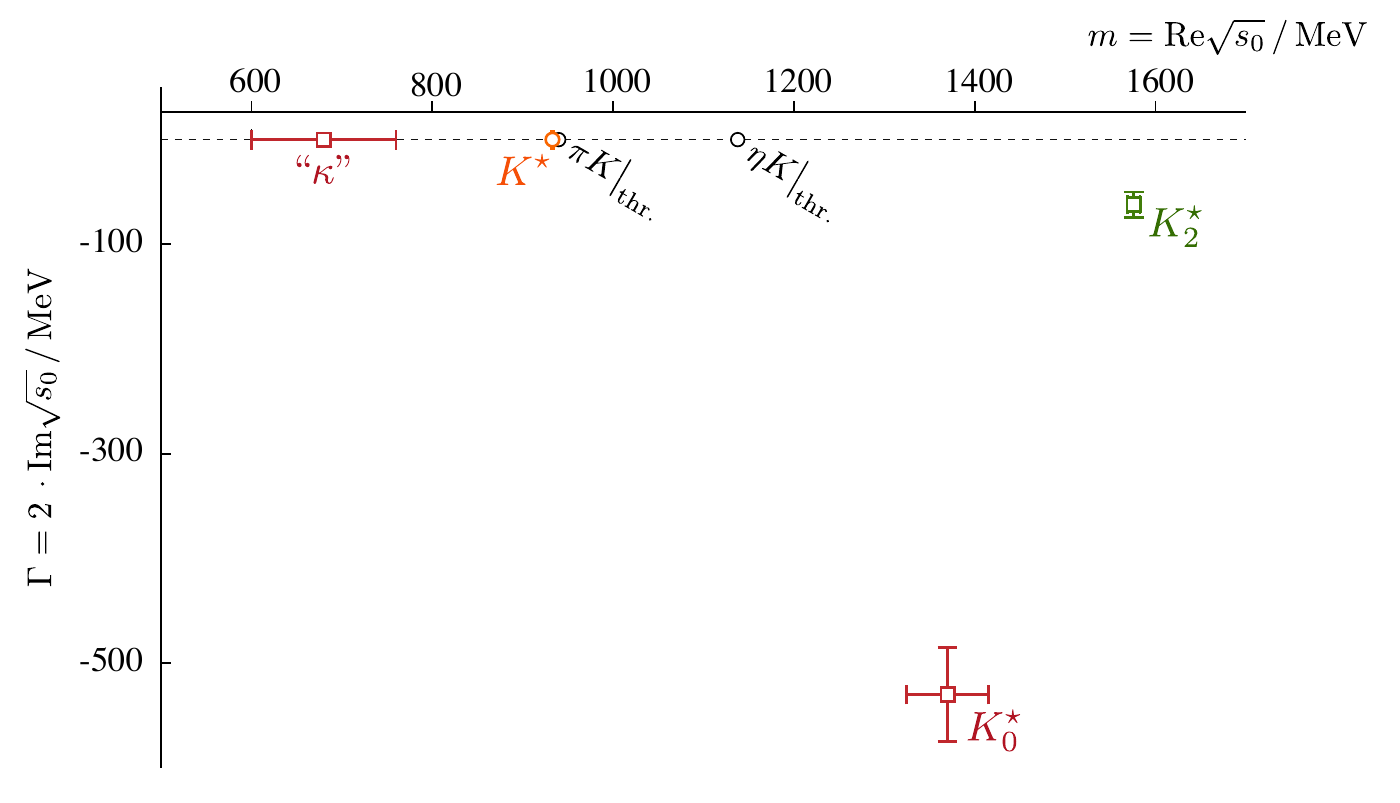}
\caption{Pole singularities of partial-wave $t$-matrices in the complex plane for $J^P=0^+$(red), $1^-$(orange) and $2^+$(green). Squares correspond to poles found on unphysical sheets, circle is a physical sheet bound-state. Uncertainties include variation under changes in parameterization form.
\label{poles} }
\end{figure}

In Fig.~\ref{poles} we present the pole singularities found for the parameterizations given previously. As expected from the relatively slow increase through $90^\circ$ in the $\pi K$ $0^+$ phase-shift, we find a $K_0^\star$ resonance pole with a large width. A pole is present in this vicinity for a range of parameterizations (not presented here) capable of describing the finite-volume spectrum, and the uncertainty on the pole position shown in Fig.~\ref{poles} includes such variation.
In the $2^+$ case the rapid rise in phase-shift is due to a $K_2^\star$ resonance pole at a slightly higher mass with a much smaller width. Both resonances have a dominant coupling to $\pi K$ extracted from the residue of the pole. The $1^-$ amplitude has a bound-state pole just below $\pi K$ threshold. The $0^+$ amplitude has another pole which lies on the real axis some way below $\pi K$ threshold. The $1^-$ pole is the only one lying on the physical sheet, the others lie on both sheet $\mathsf{II}$ where $\mathrm{Im}\, p_{\pi K} \!<\! 0$, $\mathrm{Im}\, p_{\eta K} \!>\! 0$ and sheet $\mathsf{III}$ where $\mathrm{Im}\, p_{\pi K} \!<\! 0$, $\mathrm{Im}\, p_{\eta K} \!<\! 0$. 

The extracted poles are comparable to those found in experiment; the broad scalar and narrow tensor resonances resemble the $K_0^\star(1430)$ and $K_2^\star(1430)$ experimental states respectively, although the scalar state appears to have a somewhat larger width. The heavy $(u,d)$ quark masses result in a bound-state with $J^P=1^-$ within a few MeV of $\pi K$ threshold, where experiment finds a narrow elastic scattering resonance. It is our expectation that the $K^\star$ vector bound-state will become a resonance as the light quark mass is decreased, with the observed proximity to the $\pi K$ threshold being an accident of the particular quark mass value used in this calculation. We note that the Breit-Wigner coupling extracted above, $g_R = 5.93(26)$, is in quite reasonable agreement with the value $g_R^\mathrm{phys}=5.52(16)$, corresponding to the experimental width~\cite{Beringer:1900zz}, in line with the theoretical suggestion that vector meson couplings are largely quark-mass independent~\cite{Nebreda:2010wv}.

Our observation of a $0^+$ pole below threshold on unphysical sheets agrees with the qualitative prediction of unitarized chiral perturbation theory that as the pion mass is increased above its physical value, the $\kappa$ resonance pole becomes a virtual bound state~\cite{Nebreda:2010wv}.


\paragraph{Summary:}

Through lattice QCD computation of discrete excited state spectra in finite-volumes we have been able to determine coupled-channel scattering amplitudes via parameterizations of their energy-dependence. The singularity structure of these amplitudes was then explored. This calculation serves as a demonstration that using these techniques it is possible to extract information about not only narrow resonances and bound-states, but also broad resonances and non-resonant features such as virtual bound-states. 

To compare quantitatively with experiment we would need calculations at the physical light quark mass, but the energy region of interest would then be above three-hadron (and higher multiplicity) thresholds. Formalism to relate finite-volume spectra to scattering amplitudes is not currently mature enough to be applied in that situation, although significant progress is being made~\cite{Hansen:2013dla}.

While in this case relatively little coupling was observed between the two scattering channels, there are hadron scattering situations in which strong coupling is anticipated, such as \mbox{($\pi\eta$, $\overline{K}K$)} in which the $a_0(980)$ resonance appears, and the isoscalar channel \mbox{($\pi\pi$, $\overline{K}K$, $\eta \eta$)} whose resonance structure and interpretation poses a number of questions for QCD. Building on the successful application of the finite-volume formalism for coupled-channels presented in this letter, we will consider these more strongly coupled systems, move to lighter quark masses, and explore states in exotic partial-waves.

\paragraph{Acknowledgements:} We thank our colleagues within the Hadron Spectrum Collaboration and M.R.~Pennington and A.P.~Szczepaniak for useful discussions. {\tt Chroma}~\cite{Edwards:2004sx}  and {\tt QUDA}~\cite{Babich:2010mu} were used to perform this work at Jefferson Laboratory. We acknowledge resources used at Oak Ridge Leadership Computing Facility, the Texas Advanced Computer Center and the Pittsburgh Supercomputer Center.  Support is provided by U.S. Department of Energy contract DE-AC05-06OR23177 under which Jefferson Science Associates manages Jefferson Lab, and the Early Career award contract DE-SC0006765.


\bibliography{../kpi-refs}

\end{document}